\documentclass[runningheads]{llncs}
\usepackage{graphicx}
\usepackage{subcaption}
\usepackage{multirow}
\usepackage[normalem]{ulem}
\usepackage{mathtools}
\usepackage{amsmath}
\usepackage{amsfonts}

\usepackage{xcolor}

\newcommand{\best}[1]{ \underline{\textbf{\textcolor{gray!0!black}{#1}}} }
\newcommand{\secbest}[1]{ \textbf{\textcolor{black!50!black}{#1}} }

\makeatletter
\renewcommand*{\@fnsymbol}[1]{}
\makeatother

\begin{document}
\title{Automatic Feature Extraction \\for Phonocardiogram Heartbeat Anomaly Detection using WaveNetVAE
% \footnote{Here I thank people.}
\thanks{Presented at PharML 2020 Workshop - 
\textit{European Conference on Machine Learning and Principles and Practice of Knowledge Discovery in Databases} (ECML-PKDD)}
}
\titlerunning{Automatic Feature Extraction for Heartbeat Anomaly Detection}
\author{Robert George Colt\inst{1,2}%\orcidID{0000-1111-2222-3333}
\and
Csongor Huba V\'arady\inst{1,3}
\and \\
Riccardo Volpi\inst{1}
\and 
Luigi Malag\`o\inst{1}
}

\authorrunning{R.G. Colt et al.}
\institute{
% \url{http://www.springer.com/gp/computer-science/lncs}
Romanian Institute of Science and Technology, Cluj-Napoca, Romania
\and Babes-Bolyai University, Cluj-Napoca, Romania \and
Max Planck Institute for Mathematics in the Sciences, Leipzig, Germany
 \\
\email{\{robert.colt,varady,volpi,malago\}@rist.ro}\\
}
% \accepted{PharML ECML PKDD 2020 Workshop}

%
\maketitle              % typeset the header of the contribution
\begin{abstract}
We focus on automatic feature extraction for raw audio heartbeat sounds, aimed at anomaly detection applications in healthcare.
We learn features with the help of an autoencoder composed by a 1D non-causal convolutional encoder and a WaveNet decoder trained with a modified objective based on variational inference, employing the Maximum Mean Discrepancy (MMD). Moreover we model the latent distribution using a Gaussian chain graphical model to capture temporal correlations which characterize the encoded signals. After training the autoencoder on the reconstruction task in a unsupervised manner, we
test the significance of the learned latent representations by
training an SVM to predict anomalies.
We evaluate the methods on a problem proposed by the PASCAL Classifying Heart Sounds Challenge and we compare with results in the literature.

% and we experiment with different distance and dissimilarity functions considering the temporal dimension of the latent audio representations. We compare with supervised learning techniques exploiting ad hoc feature extraction, demonstrating improvements in recall and discriminant power between anomalous and normal heartbeats.

\end{abstract}
\section{Introduction}
Anomaly detection
% aims at identifying events deviating from a normal behavior. This set of problems
is usually characterized by a class imbalance between normal and anomalous data, i.e., outliers differing from the majority of the data. In healthcare this problem is particularly relevant considering the potential for early diagnoses, by triggering expedited emergency responses in time-critical situations,
% ,which have the potential to save lives
having the potential to improve the quality of life~\cite{ukil2016iot,li2020survey,vsabichealthcare}.
% because data scarcity, complexity and heterogeneity, as well as poor mathematical characterization and non-canonical form \cite{koh2011data, cios2002uniqueness}.\\
Cardiovascular diseases are the first cause of death worldwide~\cite{li2020survey} and the problem of anomaly detection in heartbeats has been extensively approached in the literature~\cite{vsabichealthcare,ismail2018localization,krishnan2020automated,latif2018phonocardiographic}. Detecting irregularities in ECG signals can be approached efficiently with machine learning methods~\cite{vsabichealthcare,li2020survey} with considerable success,
due to the low level of noise in these signals, allowing additionally for a low sampling rate.
Classification of heartbeat anomalies from PhonoCardioGram audio signals (PCG) is a considerably more difficult task compared to ECG signals. On the other hand data are easier to obtain and successful anomaly detection algorithms based on PCG are more pervasive in the society due to the wide availability of audio recording devices.

The PASCAL Classifying Heart Sounds Challenge 2011~\cite{pascal-chsc-2011} has introduced the problem of identifying unhealthy heartbeat sounds in PCG signals. 
%in the second part of the challenge.
% , by identifying sound artifacts in otherwise normal looking heartbeats. The two datasets provided by the challenge have some of the same issues as outlined above: few samples, noise, uncharacteristic distribution of classes of real world occurrence, however this last property is helping in making the classes more balanced.\\
% The problem is introduced as a classification problem, however we can understand it through the lens of anomaly detection as the unhealthy signals are characterized by a very local-specific features like noise or extra heartbeats, in otherwise indistinguishable signals. \\
Several approaches to this problem take a fully supervised approach, by introducing expert knowledge~\cite{gomes2012classifying,deng2012robust,oliveira2014heart,avendano2010feature}, with ad-hoc features design, and by using specific wavelet transformations to identify anomalous frequencies~\cite{balili2015classification}, or through decision trees based on expert knowledge heuristics~\cite{chakir2018phonocardiogram}. Malik et al.~\cite{malik2019localization} exploited the periodicity and average heartbeat lengths to highlight anomalous behavior.

In this paper we follow a different perspective in which features are automatically extracted through an autoencoder trained on the reconstruction task.
Wang et al.~\cite{wang2016research} use CNNs and autoencoders to perform anomaly detection on time-series physiological data,  Pereira et al.~\cite{pereiraIJDMB} perform unsupervised LSTM based representation learning and anomaly detection in ECG sequences. Rushe et al.~\cite{rushe2019anomaly} introduce an anomaly detection algorithm for raw audio data, based on the ability of WaveNet~\cite{oord_WaveNet:_2016} to predict the next sample of a normal signal.
We aim at combining both approaches, in particular we would like to have a well behaved latent representation and at the same time leverage the expressivity of a WaveNet autoregressive model.
Given the importance of interpretable models in medicine~\cite{bizopoulos2018deep}, we aim at learning a set of expressive and compact features in the latent space by Variational Inference~\cite{graves2011practical}.
We demonstrate that relevant features can be automatically extracted through the reconstruction task, paving the way towards semi-supervised approaches.
% We employ a semi-supervised method to detect anomalies, by training a Variational Autoencoder in an unsupervised way to reconstruct heartbeats and clustering on their learned latent representation. The VAE uses WaveNet \cite{oord_WaveNet:_2016} as a decoder, which is composed by 1D causal convolutions with residual and skip connections and is capable of generating high fidelity natural sounds.\\
% We demonstrate comparable results to the previously mentioned state of the art methods, \csongor{and sometimes we even outperform them,} with the added bonus that our method needs no expert knowledge, and it only uses the labels to identify which cluster represents which class of points.

%%%%% METHOD section: name pending %%%%%%%
\section{Methodology}
%%%%% WaveNetAE subsection: name pending %%%%%%%
    % \subsection{WaveNetVAE}
% We develop our Variational Autoencoder based on \riccardo{"our" is not nice}
        WaveNetAE \cite{engel_neural_2017} has been proven capable of learning to reconstruct high-fidelity natural sounds like music or human speech. It consists of an encoder employing non-causal convolutional layers
        with skip connections,
        and
        % a decoder which is based
        on a conditional WaveNet~\cite{oord_WaveNet:_2016} decoder. The training is done with the objective of minimizing the negative log likelihood. Unlike simple Autoencoders, Variational AutoEncoders (VAE) \cite{kingma2013vae,rezende2014stochastic}, based on Variational Inference (VI) \cite{graves2011practical} exhibit advantageous properties by regularizing the latent space and being able to learn compact representations, by using %\csongor{[I would remove this, there is no such thing as a Graphical Independent model] a Graphical Independent model as a latent distribution i.e.}
        a multivariate Gaussian.
        
\begin{figure}[ht]
\begin{subfigure}{.5\textwidth}
  \centering
  % include second image
  \includegraphics[width=.9\linewidth]{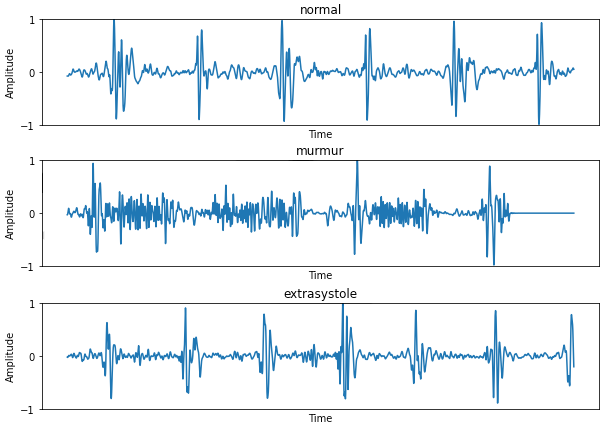}  
  \caption{Example of heartbeats signals.}
  \label{fig:HBs}
\end{subfigure}
\begin{subfigure}{.5\textwidth}
  \centering
  % include first image
  \includegraphics[width=.9\linewidth]{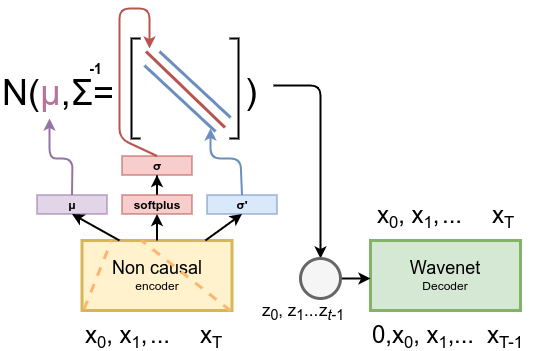}  
  \caption{Illustration of a Gaussian chain model with tridiagonal precision matrix.}
  \label{fig:WaveNetVAE}
\end{subfigure}
%\caption{Put your caption here}
\label{fig:fig}
\end{figure}

        % \robert{I would interchange the next with the second next paragraph: first talk about GC and then about MMD (which is related to both GI and GC)}
        Usually the Gaussian distribution is chosen with independent priors, i.e. with a diagonal covariance matrix.
        %In this paper we refer to this model as GI.
        In order to model a time correlation between the latent variables, we introduce a Gaussian graphical model~\cite{lauritzen1996graphical} characterized by a chain structure over the time dimension in the latent space of the encoder, cf.~\cite{apeste2017Towards}.
        %This model is referred to as GC throughout this paper.
        The overall architecture is shown in Figure~\ref{fig:WaveNetVAE} and it consists of an encoder-decoder pair. The decoder is a WaveNet model, while the distribution for the approximate posterior in the latent space is either a Gaussian Independent model (GI) or a Gaussian Chain model (GC).
        % \csongor{[I know that you wanted to explain more the difference between GI and GC, but I think you accidentally made it more confusing. Reformulate it please. I think you can just use GI after you introduce it]}
        % \robert{I think it would be better to put this paragraph after the one in which we talk about GI and before the MMD one}
        % A. Pește and L. Malagò.
        % Towards the Use of Gaussian Graphical Models in Variational Autoencoders.
        % In ICML 2017 Workshop on Implicit Models, Sydney, Australia, 10 August 2017.
        
        A known problem, when using a probabilistic distribution in the latent space of Variational Autoencoder with a powerful autoregressive decoder such as WaveNet, is that the KL term in the Evidence Lower Bound
        % Eq.~\ref{eq:ELBO})
        objective (ELBO)
        \begin{equation}
        \label{eq:ELBO}
        \mathcal{L}(x, \theta, \phi) = \mathbb{E}_{q_\theta(z|x)}[\ln p_\phi(x|z)] - D_{KL}(q_\theta(z|x) || p(z))
        \end{equation}
        might lead the optimization towards posterior collapse, as also reported in~\cite{engel_neural_2017}.
        To solve this problem we propose replacing the KL divergence in the ELBO objective with the Maximum Mean Discrepancy (MMD)~\cite{sriperumbudur2010hilbert},
        % B. K. Sriperumbudur, A. Gretton, K. Fukumizu, B. Schölkopf, and G. R. G. Lanckriet. Hilbert space embeddings and metrics on probability measures. J. Mach. Learn. Res., 11:1517–1561, 2010
        a dissimilarity measure between the aggregate posterior and the prior distribution~\cite{zhao2017infovae,tolstikhin2017wasserstein} as %seen in Eq.~\ref{eq:MMD_formula},
        %https://arxiv.org/pdf/1711.01558.pdf
        \begin{equation}
        \label{eq:MMD_formula}
        \begin{gathered}
            D_{MMD}(q||p) = \mathbb{E}_{p(z),p(z')}[k(z, z')] - 2\mathbb{E}_{q(z),p(z')}[k(z, z')] + \mathbb{E}_{q(z),q(z')}[k(z, z)], \\
        k_{gaussian}(z, z') = e^{-\frac{||z-z'||^2}{2\sigma^2}}, \qquad  k_{module}(z, z') = ||z - z'|| - ||z|| - ||z'||\;.
        \end{gathered}
    \end{equation}
    % \riccardo{the following phrase ["We found" ... "signal."] could also be removed maybe? it is more an anticipation of results?}
    % We found this in practice to reduce the posterior collapse and resulting in latent representations which carry information about the signal.
    In our experiments we used the Gaussian kernel ($k_{gaussian}$) for the GI models and for GC trained only on normal data. For the GC models trained on all both normal and anomalous data, the best results were obtained when using the module kernel ($k_{module}$) to compute the MMD.% and can also be integrated with a beta KL in the objective function, the Evidence Lower Bound (ELBO).
    % \begin{equation}
    %     \label{eq:MMD_loss}
    %     \mathcal{L}(x, \theta, \phi) = \mathbb{E}_{q_\theta(z|x)}[logp_\phi(x|z)] - D_{MMD}(q_\theta(z) || p(z))
    % \end{equation}

We train models in two ways: with all the samples (all) or only with normal samples (n).
We classify the data using a supervised SVM on the frozen latent space of the pretrained WaveNetAE (abbreviated as AE), GI and GC models which are used as feature extractors. %Secondly we perform clustering on the 3D reconstruction space (RL, MCD, PSNR).
% We consider two approaches to classify the data. Firstly we train a supervised SVM on the frozen latent space of the WaveNet Autoencoders which are then used as feature extractors. Secondly we perform clustering on the 3D reconstruction space (RL, MCD, PSNR).
% Training only on normal data the model will have worse reconstruction on the anomalous signals.
% The second approach is using all classes of data, normal and anomaly, which provides a weaker separation in the reconstruction space but the increased number of samples benefits the training algorithm in reconstruction, so it helps the classification with more accurate features. 
    % , the modified ELBO. We use an SVM classifier on the reconstruction metrics, with Radial Basis Functions in order to separate the classes. 
    
    % We use Agglomerative Clustering to cluster the latent representations of the heartbeat sounds. We experiment with different metrics like euclidean or cosine distance. However in order to better capture the temporal dimension, we use a distance based on the Fast Fourier Transform. We take the FFT of each channel of the latent representations and compute the mean squared difference between these features. The audio signal is assigned to the cluster to which it has the minimum average distance between all points inside the cluster. 
    
    % After all data points have been clustered, we use a combinatorial assignment algorithm in order to match the true label to the cluster labels.
    % At test time we classify the new heartbeats with the k-NN algorithm using the same distance. 

\section{Experimental Details}
We evaluate our methods on the Dataset B of the PASCAL Classifying Heart Sounds Challenge, including 507 records, collected with a 4,000Hz sampling frequency, and divided in three categories: \textit{Normal, Murmur} and \textit{Extrasystole}, see Fig.~\ref{fig:HBs}.
% In Figure \ref{fig:HBs} we can see examples of each type of heartbeat sound after preprocessing. Notice that on a large percentage of a signal, a normal heartbeat is indistinguishable from murmur or extrasysthole one.
Our preprocessing  consists of 3 steps: 1) we clip the signal by the $99.9$ percentile
to get rid of odd peaks;
2) we apply a low-pass filter at $195$Hz as recommended by the challenge \cite{pascal-chsc-2011} to smooth out the clipping and to remove high frequency noise; finally 3) we rescale the data between $[-1,1]$. We train on random crops of 6,144 samples.
% \robert{should we mention that this acts also as data augmentation?}.
We use WaveNet with 2 stages of 5 layers each in both the 1D non-causal Convolutional encoder and the WaveNet decoder.
% this is much reduced from the original WaveNetAE \cite{engel_neural_2017} architecture, which uses 3 stages with 10 layers each.
% The motivation for the reduction of complexity is because the frequency range of PCG signals are very reduced compared to the speech and music data, for which the WaveNetAE was designed.
For the latent space we used 4 latent channels, for each of the different models AE, GI, GC. %\riccardo{acronyms are not italics}
% with a 
%convolution of the size
Downsampling in the temporal dimension for the encoder is performed only in the final average pooling layer with stride 64 (right before the probabilistic layers for GI and GC), leading to a 1:16 compression ratio i.e. an encoding of 96 time steps with 4 channels.
We trained all models with a learning rate of $0.0002$ using Adam \cite{kingma2014adam} with default learning parameters $\beta_1=0.9$ and $\beta_2=0.999$.
We train each of the above mentioned models with a batch size of 10. Our implementation is available at  https://github.com/rist-ro/argo.
% When specified, Batch Normalization with default parameters has been used in the convolutional encoder.
% \riccardo{crops are mentioned before, channels are mentioned before, I would remove after here} Due to the high dimensionality of the data and the difference in legth of the signals, during each epoch we take only a random crop  of length 6144 from each heart beat. We use 4 channels for the latent distribution and a window of 64 time steps for downsampling the signal with max pooling, which leads to a 1:16 compression ratio. \robert{More info needed ? like number of convolution layers ?}. \csongor{it's explained above with layers and stages}

\section{Results}
In order to classify the latent representations we experiment with different crop lengths (6,144, 9,216, 12,288) and number of random crops of the raw signal. We report best results with a crop length of 12,288 and 10 crops per signal.
% Because each random crop taken from the signal could be labeled differently, 
We employ a majority voting policy between the crops of a single signal in order to assign a label.
We trained the SVM classifiers both with the raw latent representation as well as with the Fast Fourier Transforms taken on each channel separately. The best results are yielded from the latter method and are reported in Table~\ref{tab:anomaly_detection_results}. We report average on 3 training steps (spaced by 5k steps for VAEs and by 7k steps for AEs) and around the step obtaining the best classification performances on validation. We perform a 3-class classification and compute Total precision (TP) as the sum of all precisions of the 3 classes while the other metrics are computed on a binary classification task, \textit{Murmur} and \textit{Extrasystole} taken together being the positive class with \textit{Normal} as the negative class, as specified by the challenge~\cite{pascal-chsc-2011}. `\textbf{all}' specifies that the model has been trained with both anomalous and normal samples, `\textbf{n}' indicates the fact that the model was trained only with normal heartbeats. The models ending in `\textbf{bn}' use batch normalization in their encoder.
VAE models trained on all samples are more effective than models trained only on normal data, most likely due to the limited data set size. However the AE models seem to perform better when trained only on normal data obtaining good overall performances. GC models trained on all samples exhibits good overall performance as well, similarly to AE. The chain model in the temporal dimension (GC vs GI model) improves the latent space representation, which results more meaningful to the anomaly detection task. Overall, the results obtained with the proposed methods are better than those of other works dealing with this particular challenge, Table~\ref{tab:anomaly_detection_results}.

% \csongor{We should mention somewhere here that we are not as efficient in classifying in 3 classes}
\begin{table}[t]
% first column
% \begin{minipage}{0.5\linewidth}
%\begin{table}[ht]
%    \centering0
\scriptsize{
\caption{\strut \label{tab:anomaly_detection_results}
Anomaly detection for different models described in the paper, results on the test set averaged at the last 3 saved models.
% The steps of the saved models are given in parenthesis after dividing by $10^3$.
C is the regularization coefficient for a SVM using Gaussian kernel, chosen based on validation. Abbreviations: YI - Youden's Index, TP - Total Precision, Spec. - Specificity of heart problem, Sens. - Sensitivity of heart problem, DP - Discriminant Power. Second best results are highlighted in \secbest{bold} while best results are also \best{underlined}.
% \luigi{I would use bold gray and not green} \robert{You cant see gray that well} \luigi{the table is too large} \luigi{what are the numbers between parenthesis after the methods in the furst column?}
% \robert{the steps over which I took the average}
% \\\riccardo{table should be restructured..
% \begin{itemize}
%     \item too wide. Could we remove one decimal?
%     \item we should remove "new experiments" section for the final resubmission.
%     \item AUC is 0 for some experiments?
% \end{itemize}
% }
}
% \label{validation}% luigi: do not delete \label{validation} 
% \medskip
% \hskip-0.5cm
% \centering

\begin{tabular}{|c|c|c|c|c|c|c|c|}
        \hline
        Model & C & YI & TP & Spec. & Sens. & DP & AUC\\
        % \hline
        % \multicolumn{6}{|c|}{SVM Latent Representations} \\
        % Latent SVM &  &  &  &  &\\
        \hline
        
    %     AE-all (34-48) & 0.55 & 0.267 $\pm$ 0.008 & 1.541 $\pm$ 0.061 & 0.951 $\pm$ 0.019 & 0.316 $\pm$ 0.014 & 0.550 $\pm$ 0.108 & 0.682 $\pm$ 0.013 \\
        
    %     GI-all (85-95) & 0.55 & 0.235 $\pm$ 0.031 & 1.469 $\pm$ 0.041 & 0.919 $\pm$ 0.018 & 0.316 $\pm$ 0.025 & 0.402 $\pm$ 0.064 & 0.685 $\pm$ 0.003 \\
        
    %     GI-all-bn (85-95) & 0.5 & 0.292 $\pm$ 0.025 & 1.719 $\pm$ 0.083 & 0.958 $\pm$ 0.015 & 0.333 $\pm$ 0.014 & 0.603 $\pm$ 0.108 & 0.724 $\pm$ 0.010 \\
        
    %   GC-all (85-95) & 0.30 & 0.338 $\pm$ 0.021 & 2.148 $\pm$ 0.293 & 0.946 $\pm$ 0.007 & 0.392 $\pm$ 0.017 & 0.583 $\pm$ 0.045 & 0.681 $\pm$ 0.016 \\
        
    %     GC-all-bn (85-95) & 0.65 & 0.359 $\pm$ 0.007 & 1.843 $\pm$ 0.071 & 0.944 $\pm$ 0.024 & 0.415 $\pm$ 0.030 & 0.616 $\pm$ 0.086 & 0.700 $\pm$ 0.028 \\

    %     \hline
    %     AE-n (41-55) & 0.40 & 0.362 $\pm$ 0.043 & 1.883 $\pm$ 0.218 & 0.958 $\pm$ 0.007 & 0.404 $\pm$ 0.038 & 0.660 $\pm$ 0.070 & 0.721 $\pm$ 0.001 \\
        
    %     GI-n (60-70) & 0.25 & 0.281 $\pm$ 0.013 & 1.799 $\pm$ 0.042 & 0.953 $\pm$ 0.007 & 0.327 $\pm$ 0.008 & 0.553 $\pm$ 0.045 & 0.728 $\pm$ 0.011 \\
        
    %     GI-n-bn (60-70) & 0.35 & 0.224 $\pm$ 0.010 & 1.688 $\pm$ 0.054 & 0.944 $\pm$ 0.018 & 0.281 $\pm$ 0.025 & 0.461 $\pm$ 0.057 & 0.743 $\pm$ 0.006 \\
        
    %     \hline\hline
        
        AE-all & 0.55 & 0.27 $\pm$ 0.01 & 1.54 $\pm$ 0.06 & \secbest{0.95} $\pm$ 0.02 & 0.32 $\pm$ 0.01 & 0.55 $\pm$ 0.11 & 0.68 $\pm$ 0.01 \\

        GI-all & 0.55 & 0.23 $\pm$ 0.03 & 1.47 $\pm$ 0.04 & 0.92 $\pm$ 0.02 & 0.32 $\pm$ 0.03 & 0.40 $\pm$ 0.06 & 0.69 $\pm$ 0.00 \\

        GI-all-bn & 0.5 & 0.29 $\pm$ 0.03 & 1.72 $\pm$ 0.08 & \best{0.96} $\pm$ 0.01 & 0.33 $\pm$ 0.01 & 0.60 $\pm$ 0.11 & 0.72 $\pm$ 0.01 \\

       GC-all & 0.30 & \secbest{0.34} $\pm$ 0.02 & \best{2.15} $\pm$ 0.29 & \secbest{0.95} $\pm$ 0.01 & 0.39 $\pm$ 0.02 & 0.58 $\pm$ 0.04 & 0.68 $\pm$ 0.02 \\

        GC-all-bn & 0.65 & \best{0.36} $\pm$ 0.01 & 1.84 $\pm$ 0.07 & 0.94 $\pm$ 0.02 & \secbest{0.41} $\pm$ 0.03 & \secbest{0.62} $\pm$ 0.09 & 0.70 $\pm$ 0.03 \\

        \hline
        
        AE-n & 0.40 & \best{0.36} $\pm$ 0.04 & 1.88 $\pm$ 0.22 & \best{0.96} $\pm$ 0.01 & 0.40 $\pm$ 0.04 & \best{0.66} $\pm$ 0.07 & 0.72 $\pm$ 0.00 \\

        GI-n & 0.25 & 0.28 $\pm$ 0.01 & 1.80 $\pm$ 0.04 & \secbest{0.95} $\pm$ 0.01 & 0.33 $\pm$ 0.01 & 0.55 $\pm$ 0.04 & \secbest{0.73} $\pm$ 0.01 \\

        % GI-n-bn &0.35 & 0.22 $\pm$ 0.01 & 1.69 $\pm$ 0.05 & 0.94 $\pm$ 0.02 & 0.28 $\pm$ 0.03 & 0.46 $\pm$ 0.06 & \best{0.74} $\pm$ 0.01 \\
        
        GC-n & 0.10 & 0.29 $\pm$ 0.03 & 1.87 $\pm$ 0.46 & \secbest{0.95} $\pm$ 0.00 & 0.33 $\pm$ 0.02 & 0.56 $\pm$ 0.04 & \secbest{0.73} $\pm$ 0.00 \\

        % % \hline
        % % \multicolumn{6}{|c|}{SVM in the Reconstruction Space} \\
        % \hline
        % \multirow{2}{*}{SVM Rec}
        %                           &     GI     & 0.12 & 1.1 & 0.47 & 0.65 & 0.12\\
        %                           &    GC     &  &  &  &  & \\
        % % \hline
        
        % % \multicolumn{6}{|c|}{Clustering in the Reconstruction Space} \\
        % \hline
        % \multirow{2}{*}{Clustering Rec}
        %                         &    GI     &  &  &  &  & \\
        %                         &    GC     &  &  &  &  & \\
        \hline\hline
        %Balili et. al. 2015 # luigi: we need to save space
        \multicolumn{2}{|c|}{Balili et al.~\cite{balili2015classification}} & 0.15 & 1.36 & \secbest{0.95} & 0.2 & 0.37 & N/A\\
        %Chakir et. al. 2018 # luigi: we need to save space
        \multicolumn{2}{|c|}{Chakir et al.~\cite{chakir2018phonocardiogram}} & 0.15 & 1.58 & 0.66 & \best{0.49} & 0.15  & N/A\\
        \multicolumn{2}{|c|}{Zhang et al.~\cite{zhang2017heart}} & 0.29 & \secbest{2.03} & \secbest{0.95} & 0.34 & 0.54  & N/A\\ %column SVM-DM
        % Malik et. al. 2019 \cite{malik2019localization} & - & - & - & 0.98? & -\\ Csongor - these numbers are bullshit IMHO # luigi: good
        \hline
    \end{tabular}
\hrule height 0pt
%\end{table}
% \end{minipage}
%second column
% \begin{minipage}{0.5\linewidth}
% \centering
% \includegraphics[width=0.4\linewidth]{images/3x3_rec_test.png}
% \captionof{figure}{\strut TODO \luigi{This is placeholdeer until the  figue is final} \csongor{What figure should come here?} \robert{right now this figure looks preety bad, also on train and validation...} \label{x}}% luiggi do note delete \label{x}
%     \label{fig:rec_3x3}
% \end{minipage}
}
\end{table}

\section{Conclusions}
We demonstrate how relevant features for PCG audio signals can automatically be extracted through WaveNet autoencoders.
% , and how a Variational Wavenet Autoencoders are less prone to overfitting in the task considered.
% , paving the way towards semi-supervised approaches.
We introduce a WaveNetVAE model, trained using MMD in the latent space and we demonstrate how the introduced regularization produce a benefit in terms of SVM classification in the latent space. Additionally we found that Batch Normalization in the encoder produce benefits in terms of latent representations for the WaveNetVAE models.
% show how leveraging the latent space representations of the WaveNet autoencoders.
We obtained better results than other works dealing with the PASCAL Classifying Heart Sounds Challenge 2011, evaluated with several metrics of interest for the challenge. We show how a VAE or AE model can be used to automatically extract relevant features to the anomaly detection task, without the need of expert domain knowledge. We chose a simple method to classify the frozen latent space of the heartbeats (SVM), to probe the latent space representation learned by the autoencoders. The approach presented paves the way towards semi-supervised/self-supervised training for detecting anomalies in audio signals.

% \luigi{can you check the arxiv papers in the bib if they have been published somewhere recently? I am referring to wavenet for instance.}
% \robert{Wavenet is only on arxiv. The others too. For instance wavenet: 
% https://research.google/pubs/pub45774/ }

\section{Acknowledgements}
This work was supported by the DeepRiemann project, co-funded
by the European Regional Development Fund and the Romanian Government through the Competitiveness Operational Program 2014-2020, Action 1.1.4, project ID P\_37\_714, contract no. 136/27.09.2016.

% \csongor{Good question, if we add acknowledgements, we should also link to the github}
% \robert{i remember Larisa said once to add the acknoledgements.}
% \riccardo{yes, we should add acknowledgements, in the same fashion as:
% "This work was supported by the DeepRiemann project, co-funded
% by the European Regional Development Fund and the Romanian Government through the
% Competitiveness Operational Program 2014-2020, Action 1.1.4, project ID P\_37\_714,
% contract no. 136/27.09.2016."
\bibliographystyle{splncs04}
\bibliography{bibliography}
%





% LNCS Homepage, \url{http://www.springer.com/lncs}. Last accessed 4
% \end{thebibliography}
\end{document}